\documentclass[%
 aip,
 amsmath,amssymb,
 reprint,%
]{revtex4-1}

\usepackage{graphicx}
\usepackage{dcolumn}
\usepackage{bm}

\usepackage[utf8]{inputenc}
\usepackage[T1]{fontenc}
\usepackage{mathptmx}
\usepackage{etoolbox}
\usepackage{comment}

\makeatletter
\def\@email#1#2{%
 \endgroup
 \patchcmd{\titleblock@produce}
  {\frontmatter@RRAPformat}
  {\frontmatter@RRAPformat{\produce@RRAP{*#1\href{mailto:#2}{#2}}}\frontmatter@RRAPformat}
  {}{}
}%
\makeatother
\begin{document}

\preprint{AIP/123-QED}

\title{Energy conversion and scaling analysis of relativistic magnetic reconnection
}
\author{Harihar Pradhan}
\email{ph22resch11004@iith.ac.in}
\author{Kirit D Makwana}%
\affiliation{ 
Department of Physics, Indian Institute of Technology, Hyderabad, Sangareddy, Telangana 502284, INDIA
}%

\author{Bart Ripperda}
\affiliation{%
Department of Physics, University of Toronto, 60 St. George Street, Toronto, ON M5S 1A7, Canada
}%
\affiliation{David A. Dunlap Department of Astronomy, University of Toronto, 50 St. George Street, Toronto, ON M5S 3H4, Canada}
\affiliation{Perimeter Institute for Theoretical Physics, 31 Caroline St. North, Waterloo, ON N2L 2Y5, Canada}

\date{\today}

\begin{abstract}
Relativistic magnetic reconnection is a key process for accelerating charged particles and producing high-energy radiation. We study this process using relativistic resistive magnetohydrodynamics simulations. Starting with Harris sheet configuration, we study time evolution of reconnection rate and the Alfv\'en four Mach number for outflow.  These measurements validate the Sweet-Parker scaling, consistent with previous studies. To study energy conversion processes, we calculate Ohmic dissipation, crucial for understanding how energy is converted between plasma and electromagnetic fields. Decomposing electric field components relative to velocity field, we find that energy conversion is initially dominated by the resistive electric field, but convective electric fields take over as reconnection progresses. Plasma primarily gains energy within the current sheet and near the separatrix. We perform a scan of magnetization for mildly relativistic plasma to examine scaling laws previously derived for non-relativistic inflow. We find the inflow is slower than predicted, due to conversion of magnetic energy mostly into thermal energy, causing strong compressibility. We calculate and verify the scaling of the compressibility factor, providing a more accurate representation of inflow dynamics. We analyze the impact of a guide field on reconnection and energy partition, finding that a stronger guide field reduces the reconnection rate but has minimal effect on the relative distribution of kinetic, magnetic, and thermal energy. Addition of rotating guide field and variations in initial pressure and density have little effect on the energy composition of the outflow, with thermal energy consistently dominating at nearly 90\%.
\end{abstract}
\maketitle
\section{\label{sec:Intro}Introduction}
Magnetic reconnection is a fundamental process in which magnetic field lines break, reconnect, and release stored magnetic energy, transforming it into the kinetic and thermal energy of the plasma. This phenomenon is crucial in the study of space, laboratory, and astrophysical plasmas\cite{1986Biskamp, Priest2011}, as it provides key insights into energy conversion mechanisms in high-energy plasma environments. Relativistic magnetic reconnection occurs in highly magnetized and high energy environments where the magnetic energy density becomes comparable to rest mass energy density of plasma\cite{Blackman,Tenbarge,Tolstykh}. It is particularly significant in understanding the dynamic processes occurring in astrophysical compact objects, such as soft gamma-ray repeaters\cite{Lyutikov2006,GillHeyl2010,Masada2010},pulsar magnetosphere\cite{Arons2012,Uzdensky_2014,Philippov2022}, pulsar winds\cite{2001ApJ...547..437L,Arons2012}, active galactic nuclei\cite{DiMatteo1998,Giannios2010,Sironi_2021,Ripperda_2022}, and gamma-ray bursts\cite{Drenkhahn2002,McKinney2011,Zhang_2011}.

The rapid reconfiguration of magnetic field connectivity facilitates the conversion of magnetic energy stored in antiparallel magnetic fields into plasma energy\cite{Sweet1958,Parker1957,Parker1963,Birn2012}, enabling efficient particle acceleration and heating\cite{Drake2006,Li2019}. Recent general relativistic magnetohydrodynamics (GRMHD) simulations\cite{Bart2020} have demonstrated that magnetic reconnection can produce magnetic islands and provide a plausible explanation for black hole flares\cite{Ripperda_2022}. To investigate the mechanisms and regions responsible for this energy conversion, several recent studies have analyzed the statistical properties of the energy transfer rate per unit volume from electromagnetic energy to plasma energy, represented by Ohmic dissipation\cite{Yang2016,Matthaeus2020,Du_2022,Pongkitiwanichakul_2021}. These studies, primarily focused on non-relativistic or weakly collisional plasmas, and their direct applicability to relativistic reconnection remains uncertain, as the energy conversion mechanisms and plasma conditions in relativistic regimes may differ significantly.
\\
Magnetohydrodynamics (MHD) is crucial for understanding large-scale magnetic reconnection in astrophysical systems, particularly in environments that are collisional. In such astrophysical environments, like the warm ionized interstellar medium (ISM) and galaxy clusters, where the mean free path, $\lambda_{mf}$ is small compared to the size of the system $L$ (the ratio $\lambda_{mf}/L$ is of the order of $10^{-8}$  for warm ionized ISM and $10^{-2}$ for galaxy clusters)\cite{Schekochihin_2009}. These regions exhibit collisional plasma, where MHD becomes an essential framework. Relativistic systems, such as accretion disks for luminous active galactic nuclei (AGN), may be weakly collisional. In such regimes, the relativistic resistive MHD (RRMHD) framework, free from kinetic scales such as the plasma skin depth and gyro radius, is particularly suited for exploring reconnection processes in high-energy systems at large scales. Many relativistic systems, such as pulsar wind nebulae (PWN) and extragalactic jets, are effectively collisionless, with mean free path $\lambda$ much larger than system sizes L ($\lambda>>L$), making an MHD description less applicable and necessitating a kinetic approach like particle-in-cell (PIC) simulations. However, astrophysical systems often exhibit extreme scale separations, where system sizes exceed kinetic scales by factors of $10^9 - 10^{17}$. Resolving these scales with PIC simulations is computationally infeasible\cite{Guo_2020}. While PIC method provides valuable insights into kinetic processes, extrapolating their conclusions to larger scales is uncertain. Previous studies\cite{10.1063/1.5037774,Makwana_2018} have demonstrated that coupled MHD-PIC simulations can capture kinetic effects while reducing computational costs, making them suitable for collisionless reconnection scenarios. However, for collisional and large-scale plasmas, where kinetic effects are less dominant, fully MHD simulations provide a more computationally efficient and effective approach. Also the MHD-PIC coupling is yet to be developed for the relativistic regime. This underscores the need for models like MHD, which effectively capture large-scale dynamics while subgrid effects are described with a resistivity.

In the study of relativistic magnetic reconnection, the high sigma ($\sigma\equiv B^2/\rho\Gamma^2$) regions—where the magnetic energy density greatly exceeds the plasma rest mass energy density remain largely unexplored. Previous studies of relativistic reconnection have explored scaling laws up to moderately high$-\sigma$ \cite{guo2024magneticreconnectionassociatedparticle}. The range of $\sigma = 1 - 100$ remains largely unexplored due to the difficulty of accurately describing high$-\sigma$ regions for conventional MHD methods. Moreover, the mechanisms governing energy conversion in these extreme environments are still not fully understood.
In this paper, we conduct 2.5D MHD simulations to study plasma dynamics and energetics at varying magnetic Reynolds numbers and magnetization parameters. In Section \ref{sp-scale}, we validate the Sweet-Parker scaling. In Section \ref{Erg_conserv}, we analyze the energy conversion processes during reconnection. In Section \ref{sigma-scale}, we perform a parameter scan over $\sigma$ to identify discrepancies with previously proposed scaling laws. In Section \ref{Erg-Part}, we investigate the energy partitioning near the reconnection region. In Section \ref{Guide-field}, we examine the influence of a guide field on plasma dynamics and energy distribution.


\section{Simulation setup}
We conducted 2.5D special relativistic resistive magnetohydrodynamic (SRRMHD) simulations using the BHAC (Black Hole Accretion Code)\cite{BHAC,Olivares2019,Ripperda_2019} by considering a flat Minkowski metric. Our simulation setup closely follows the configuration described in [\onlinecite{takahashi}]. We initialize the system using a Harris current sheet\cite{Harris1962} in a cartesian (x,y,z) coordinate system. The magnetic field has the form
\begin{equation}
    B_y=B_0\tanh(2x/\lambda),
    \label{eq:tanh}
\end{equation}

where $B_0$ is the magnitude of the magnetic field far away from reconnection zone, $\lambda=0.02$ is the thickness of the current sheet, and we consider the total box length to be unity ($l=1$). The pressure profile is $P=P_0 + 0.5*[B_0^2-B_y^2]$, and density is $\rho=P/T_0$ where $P_0$ and $T_0$ are initial gas pressure and temperature respectively, chosen as $P_0=1,\quad T_0=1$. Here, density is in rest frame, whereas the rest of the quantities ($\textbf{u},\,\textbf{B},\,\text{P}$) are in the lab frame. The adiabatic index for the relativistic plasma is set to $\gamma=4/3$. To start the magnetic reconnection, we introduce a perturbation given in the form of magnetic vector potential as $\delta A_Z=-\delta B_0\lambda*exp[-(x^2+y^2)/\lambda^2]$, here $\delta B_0=0.03B_0$ taken to efficiently initiate reconnection. 
\\
Plasma $\beta_m$ is defined as $\beta_m=2P_0/B_0^2,$ and the magnetization parameters are defined as $\sigma\equiv B^2/\rho\Gamma^2$ and $\sigma_H \equiv B^2/(\rho + \frac{\gamma}{\gamma-1}P)$, where $\Gamma=1/\sqrt{1-v^2/c^2}$ is the Lorentz factor and the flow velocity $v$ is taken in a region far from current sheet. We have considered relativistic Alfven velocity ($v_A=\frac{B}{\sqrt{\rho+\frac{\gamma}{\gamma-1}P+B^2}}$) as derived in [\onlinecite{Keppens}] giving the Alfven four speed as $u_A=\Gamma v_A$. Magnetic Reynolds number is defined as $R_M=\lambda/\eta$, where $\eta$ is the resistivity\cite{takahashi}. We set speed of light $c=1$ and normalize 4$\pi$ to unity.\\
Time is normalized by the Alfven crossing time $\tau\equiv \lambda/v_A$. We considered continuous boundary conditions on all sides with boundaries at $x,y=\pm 0.5$. The simulation domain has grid resolution up to $4096\times 4096$ cells. The resolution is systematically increased until the results converge, ensuring that the effects of numerical resistivity were minimized in the simulation. This approach guarantees that the observed reconnection dynamics are governed by the physical resistivity introduced in the model rather than by artifacts arising from numerical discretization, as confirmed by the convergence of the reconnection rate with increasing grid resolution in Section \ref{sp-scale}.
 

\section{Results}
\subsection{\label{sp-scale}Sweet-Parker Scaling}
\begin{figure*}
\includegraphics[scale=0.75]{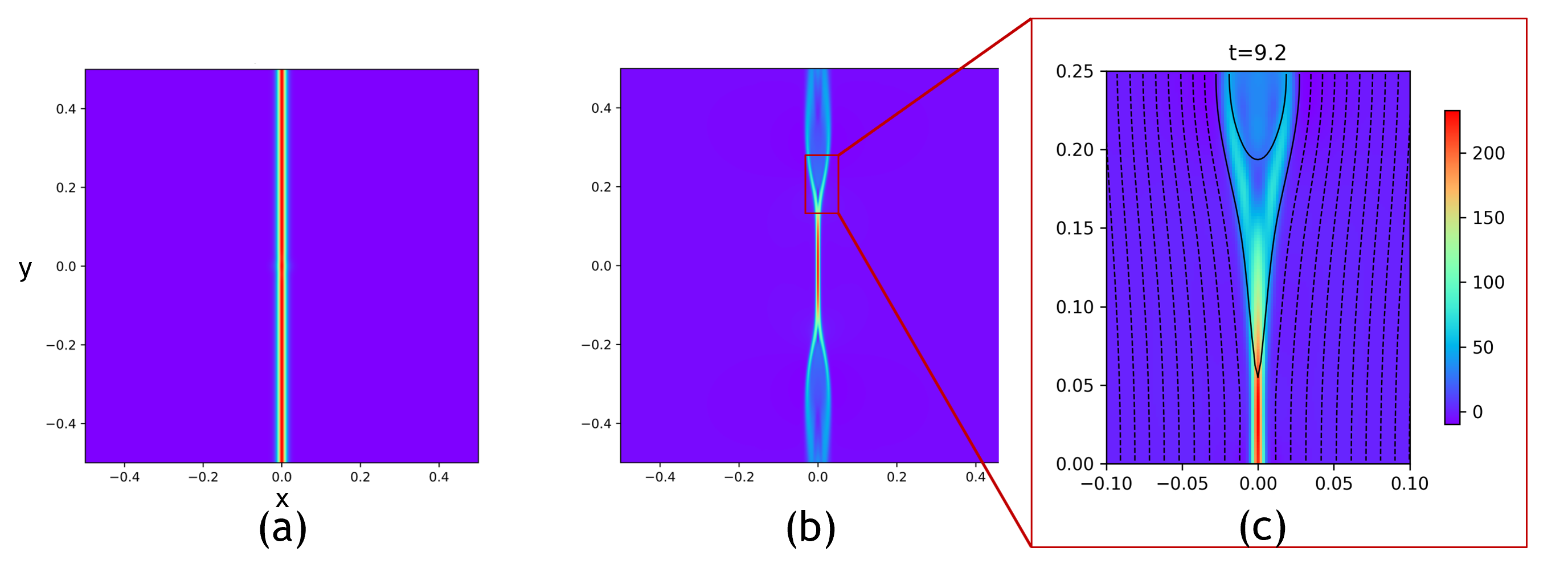}
\caption{\label{fig:Plasmoid}Snapshots of z component of current density ($J_z$) taken at time (a) $t/\tau=0$ showing initial current sheet without perturbation and (b) $t/\tau=155$ shows the magnetic islands formed during reconnection. Section of a magnetic island is zoomed in (c). Black lines represent the magnetic field lines.}
\end{figure*}

The current sheet formed during the simulation in Fig.~\ref{fig:Plasmoid}(a) gets squeezed in because of the perturbation that initiates the reconnection. This process drives plasma inflow into the current sheet in the $x$-direction near the reconnection region. As reconnection progresses, magnetic islands are formed within the current sheet due to the tearing instability, as shown in Fig.~\ref{fig:Plasmoid}(b - c). Subsequently, these are ejected in the $y$-direction, leaving the simulation domain.

\begin{figure*}
\includegraphics[scale=0.3]{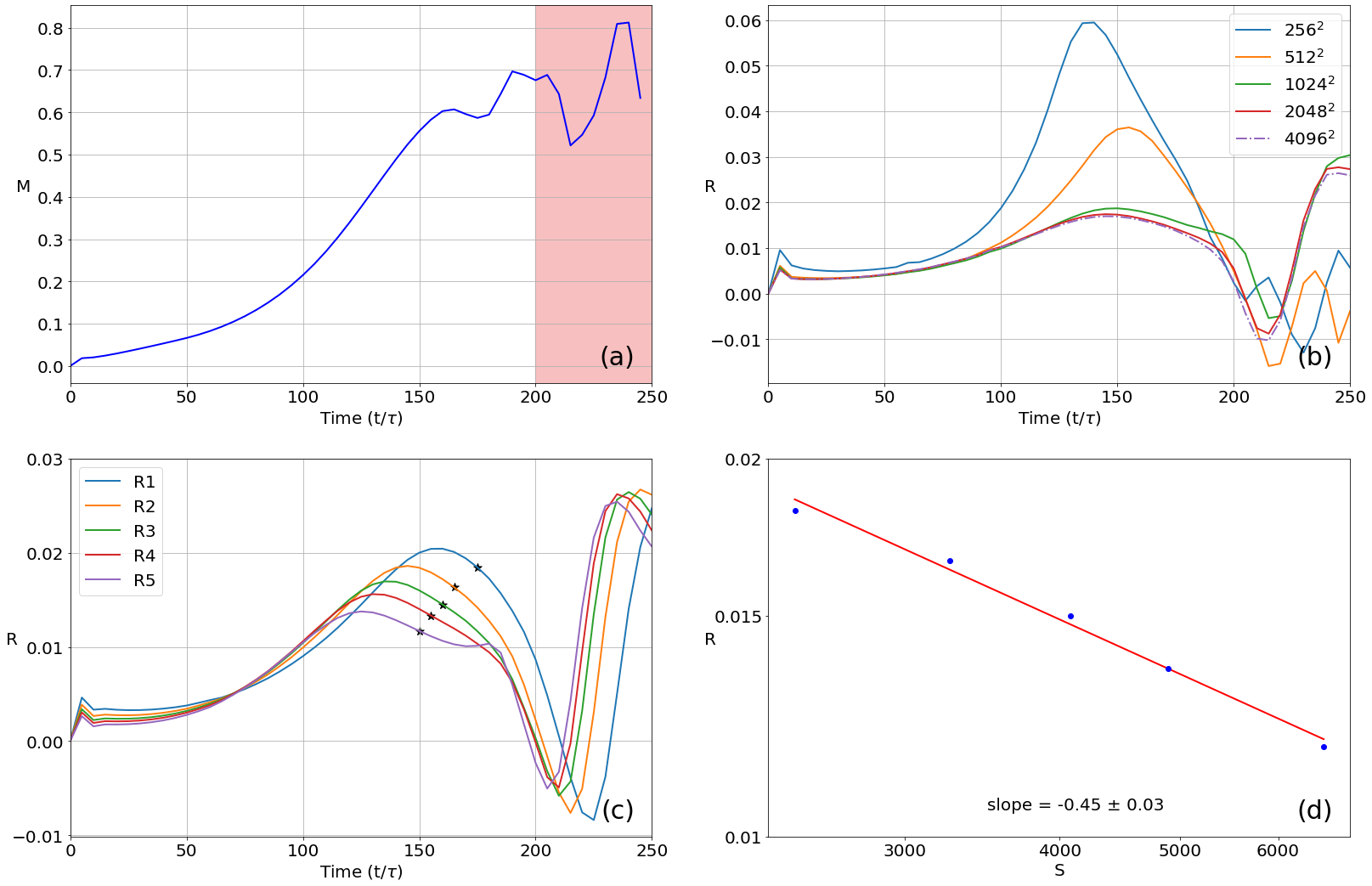}
\caption{\label{fig:reccon} Panel (a) shows temporal evolution of Alfven four Mach number along the $x=0$ axis; where the red region highlights the time when the magnetic island has exited the domain. The evolution saturates at $t/\tau = 155 $. Panel (b) demonstrates the convergence of results at high resolution of $2048^2$. Panel (c) depicts the reconnection rate for different Reynolds number as shown in Table~\ref{tab: Recc_rate}, where markers indicate the time at which the reconnection rate is measured, at $(t/\tau )_{20}$. Panel (d) illustrates the scaling of reconnection rate with Lundquist number $S=R_ML_{20}v_A/\lambda$ consistent with the relation $R\sim S^{-0.5}$.}
\end{figure*}

To characterize the reconnection outflow, we analyze the evolution of the maximum Alfven four Mach number, $M_A$ which is defined as the ratio of the maximum outflow four-velocity along sheet length ($x=0$) to the Alfven four-speed, $M_A=u_{y,max}/u_A$. Fig.~\ref{fig:reccon}(a) illustrates the time evolution of $M_A$. Here, the outflow experiences an initial acceleration when the reconnection begins ($0\leq t/\tau\leq 150$) and it gradually slows down when the reconnection region is completely formed ($150\leq t/\tau \leq 200$) because of the collison of outflow with the magnetic islands. The outflow again accelerates when the magnetic island leaves the simulated region ($t/\tau \geq 200$).

Fig.~\ref{fig:reccon}(b) shows the convergence of results at $2048^2$. Our analysis was done for different resolutions [25$6^2$, 51$2^2$, 102$4^2$, 204$8^2$, 409$6^2$] to ensure convergence of reconnection rate. Fig.~\ref{fig:reccon}(c) represents the time evolution of reconnection rate $R=|v_x|/v_A$, where $v_x$ is calculated at $x=\pm 0.05$ (close to edge of current sheet) by taking an average over a line $y \in[-0.2,\,0.2]$. Due to symmetry, the values on both sides of current sheet are nearly equal. We observe that the reconnection rate in R3 increases to a maximum value ($\sim 0.017$), showing the initial acceleration of plasma inflow. However, as thermal pressure begins to build up in the current sheet, the reconnection rate subsequently decreases ($t/\tau\geq 135$). This thermal feedback arises because the accumulated pressure resists further compression of the plasma and reduces the inflow velocity into the reconnection region.

\begin{table}
\caption{\label{tab: Recc_rate}
Parameter ($\beta_m=0.2$, $\lambda=0.02$)}
\begin{ruledtabular}
\begin{tabular}{cccc}
ID&$R_M$&Reconnection rate(R)$ (10^{-2})$&($t/\tau)_{20} $\\
\hline
R1&300& 1.82& 175 \\
R2&400& 1.66& 165  \\
R3&500& 1.5& 160 \\
R4&600& 1.36& 155 \\
R5&800& 1.18& 150 \\
\end{tabular}
{\raggedright \textbf{Notes:} Magnetic Reynolds number ($R_M=\lambda/\eta$); Column 4: time at which current sheet length becomes $0.4$.  \par }
\end{ruledtabular}
\end{table}

To determine the scaling behaviour of the reconnection rate R, we vary the magnetic Reynolds number $R_M$ over a range of values (300 - 800) by varying the resistivity ($\eta$). At each value of $R_M$ the reconnection rate is measured at time $(t/\tau)_{20}$ as shown in Table~\ref{tab: Recc_rate}, where $(t/\tau)_{20}$ corresponding to time when the current density ($J_{0.2}$) at position $L_{20}\ [0,\,0.2]$ is reduced to half of the current density ($J_{0}$) at the centre of the sheet $L_0\ [0,\ 0]$, i.e $J_{0.2}/J_0=0.5$ similar to [\onlinecite{takahashi}]. At this stage, the reconnection process can be considered to be in a quasi-steady state, as the outflow velocity approaches saturation before the magnetic island exits the simulation domain. This procedure ensures consistency in measuring the reconnection rate across different cases. This was repeated by varying the Reynolds number $R_M=[300,\ 400,\ 500,\ 600,\ 800]$ and checking the convergence for each run. The results of this analysis are summarized in Fig.~\ref{fig:reccon}(c), which shows the dependence of the reconnection rate on the magnetic Reynolds number. Where, the maximum reconnection rate and the time taken to reach $(t/\tau)_{20}$ decreases with increase in magnetic Reynolds number. Markers are located at time $(t/\tau)_{20}$ where the reconnection rate is taken for comparison. These reconnection rates are plotted against their corresponding magnetic Reynolds number in Fig.~\ref{fig:reccon}(d) which gives a scaling of $R\simeq S^{-0.45\ \pm\  0.03} $, confirming that the obtained reconnection rate follows a scaling relation consistent with the Sweet-Parker theory, where $R\propto S^{-0.5}$ is obtained in [\onlinecite{Lyubarsky05}].
   
\subsection{\label{Erg_conserv}Energy conversion}
\begin{figure}
\includegraphics[scale=0.25]{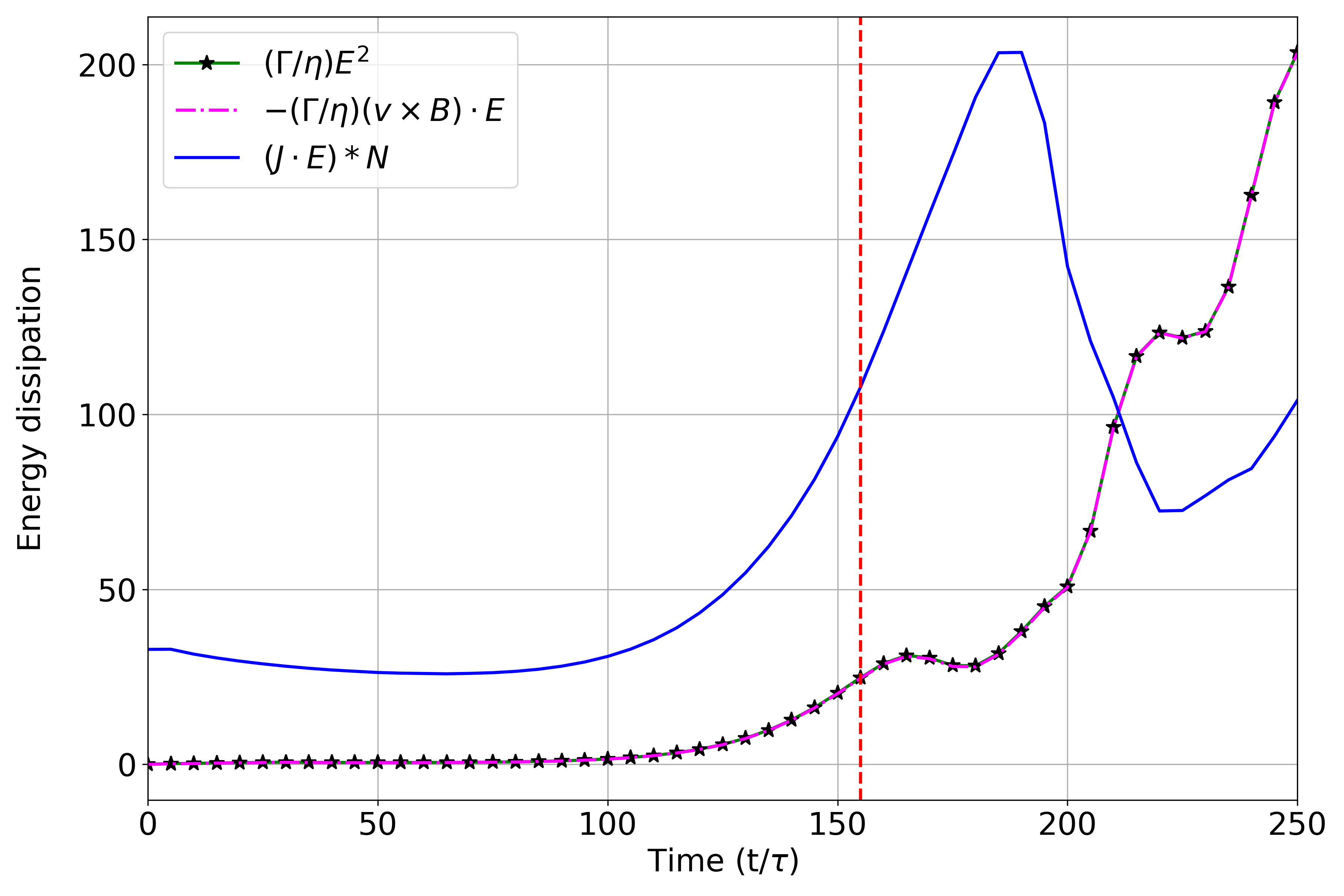}
\caption{\label{fig:JE1}The time evolution plot shows that the small difference between the two terms (green and magenta) on RHS of Eq.~(\ref{eq:J.E}) is responsible for the energy conversion term. This, $\textbf{J}\cdot\textbf{E}$ (blue) is the difference between greeen and magenta line multiplied by factor $N=617$ to make the small value visible and compare its evolution with other terms. The red dashed line indicates the time $(t/\tau)_{20}$. Energy conversion increases as the simulation progresses and dips when the island exits the domain.}
\end{figure}

In order to understand the energy conversion first we obtain the $\textbf{J}\cdot \textbf{E}$ term, describing Ohmic dissipation. Taking dot product of $\textbf{B}$ with Faraday's law, we get
\begin{equation}
    \frac{\partial}{\partial t}\Big{(}\frac{\text{B}^2}{2}\Big{)} = \boldsymbol{\nabla}\cdot(\textbf{B}\times\textbf{E})-(\boldsymbol{\nabla}\times\textbf{B})\cdot\textbf{E}
\end{equation}
Using Ampere's law $\boldsymbol{\nabla}\times \textbf{B}=\textbf{J}+\frac{\partial\textbf{E}}{\partial t}$ we get,
\begin{equation}
    \frac{\partial u_{EM}}{\partial t} + \nabla\cdot \textbf{S}=-\textbf{J}\cdot \textbf{E}
\end{equation}
where,    $u_{EM}=\frac{B^2+E^2}{2},\textbf{  } \textbf{S}=\textbf{E}\times \textbf{B}$.
The $\textbf{J}\cdot \textbf{E}$ acts as a source term and measures the energy conversion from electromagnetic energy to plasma energy.
\\
We use the relativistic Ohm's law\cite{Komissarov_2007,Blackman&Field_1993},
\begin{equation}
    \textbf{J} = q\textbf{v} + \frac{\Gamma}{\eta}[\textbf{E} + \textbf{v}\times \textbf{B}-(\textbf{E}\cdot \textbf{v})\textbf{v}]
    \label{eq:relv_ohms_law}
\end{equation}
where $q=\nabla\cdot \textbf{E}$. Since there is no component of the electric field ($\text{E}_x,\ \text{E}_y$) in the direction of the velocity field throughout the simulation; we get,
\begin{equation}
    \textbf{J}\cdot \textbf{E} = \frac{\Gamma}{\eta}[\text{E}^2 +(\textbf{v}\times \textbf{B})\cdot\textbf{E}]
    \label{eq:J.E}
\end{equation}
The terms of Eq.~(\ref{eq:J.E}) are summed over the full box and are plotted in Fig.~\ref{fig:JE1}. Both the terms on the right-hand side (RHS) are comparable in magnitude, and the small difference between them plays a crucial role in driving the energy conversion process. This small difference corresponds to the non-ideal electric field, since the total electric field can be decomposed as $\textbf{E}=\textbf{E}_\text{ideal} + \textbf{E}_\text{non-ideal}=-\textbf{v} \times \textbf{B} + \textbf{E}_\text{non-ideal}$, implying that $\textbf{E}_\text{non-ideal}=\textbf{E} + \textbf{v}\times \textbf{B}$. This difference is responsible for transferring energy from the electromagnetic field to the plasma during magnetic reconnection. We observe that the energy conversion keeps on increasing as the reconnection progresses and then, there is a rapid fall in its value, once the lower region of the magnetic island moves out of the simulation domain. The $\textbf{J}\cdot\textbf{E}$ term shown here is calculated in two ways, first by directly doing a dot product of current density and electric field vectors and secondly, by adding up the RHS terms of Eq.~(\ref{eq:J.E}), and both the values consistently match each other. Here the electric field and current density are perpendicular to the plane of the velocity field. Decomposing the components of the electric field with respect to velocity leaves us only with the perpendicular $\textbf{J}\cdot \textbf{E}_{\perp,v}$ term, where $\textbf{J}\cdot \textbf{E}_{\perp,v}=\textbf{J}\cdot\textbf{E}-\textbf{J}\cdot \textbf{E}_{\parallel,v}$, and $\textbf{E}_{\parallel,v}$ is the component of the electric field along the direction of velocity. Hence from Eq.~(\ref{eq:relv_ohms_law}), we have,

\begin{equation}
    (\textbf{J}\times \textbf{v})\times \textbf{v} = \frac{\Gamma}{\eta}[(\textbf{E}\times \textbf{v})\times \textbf{v} - [\textbf{v}(\textbf{v}\cdot \textbf{B})]\times \textbf{v} +(\text{v}^2\textbf{B})\times \textbf{v}]
\end{equation}
\begin{equation}
    -\textbf{J}\text{v}^2 = (\Gamma/\eta)[-\text{v}^2\textbf{E}+(\text{v}^2\textbf{B})\times\textbf{v}]
    \label{eq:Jvsq}
\end{equation}
\begin{equation}
    \textbf{J}\cdot \textbf{E}_{\perp,v} = -\textbf{J}\cdot(\textbf{v}\times \textbf{B})+(\eta/\Gamma)\text{J}^2
    \label{eq:J.Eperp}
\end{equation}
 Eq.~(\ref{eq:Jvsq}) comes from the fact that current density ($\text{J}_z$) and electric field ($\text{E}_z$) components are only non-zero perpendicular to the velocity field in the simulation. Taking a dot product on both sides with current density gives us Eq.~(\ref{eq:J.Eperp}), which we use to analyze the temporal and spatial evolution of energy transfer. This will aid in understanding the energy conversion processes during magnetic reconnection. 

\begin{figure}
\includegraphics[scale=0.55]{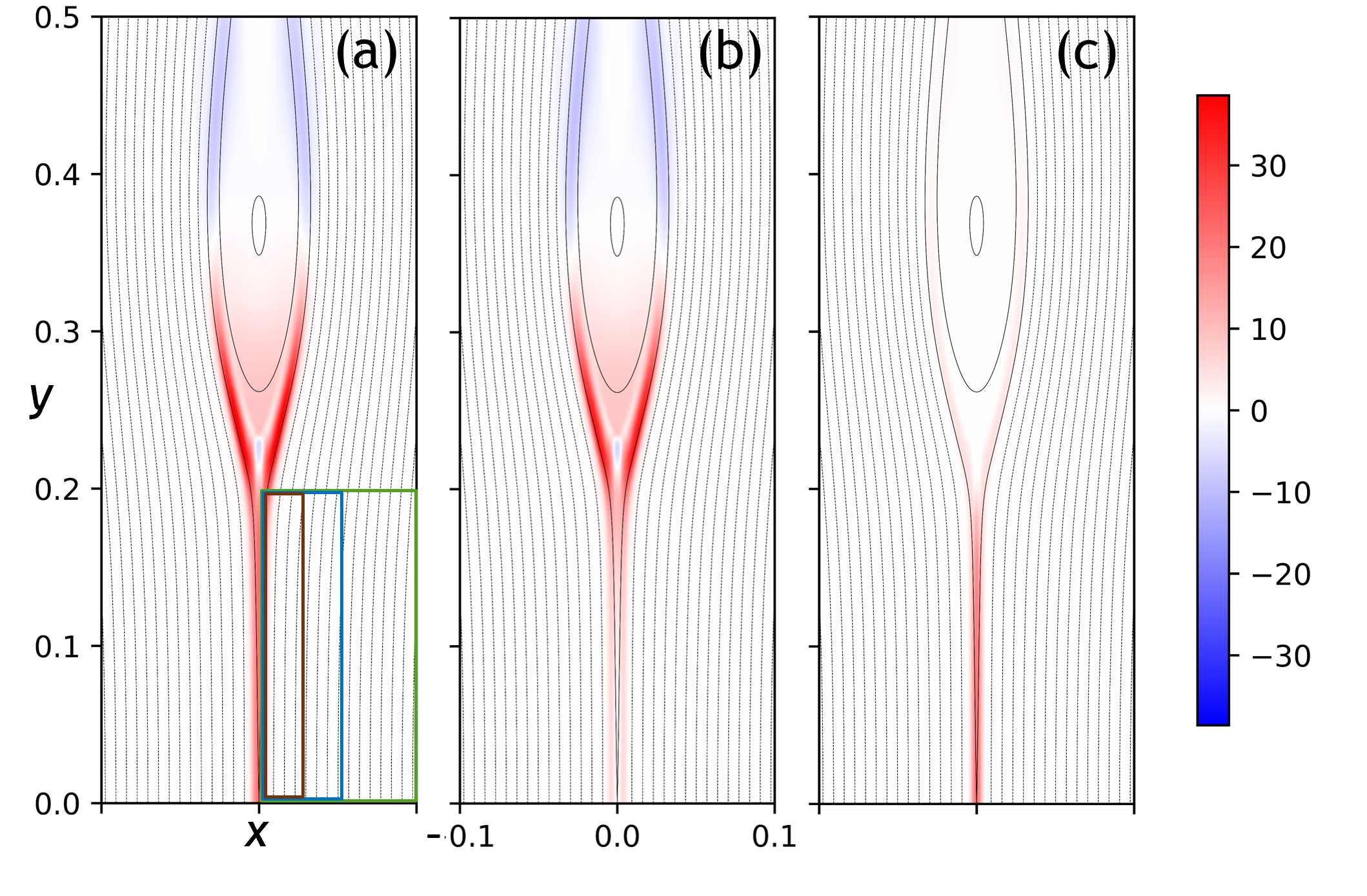}
\caption{\label{fig:Conv_res_spatial} Spatial distribution of convective and resistive electric field, showing the regions of energy conversion. Blue region indicates the conversion of plasma energy to electromagnetic energy, whereas the red regions show the vice-versa. The three panels show the terms of Eq.~(\ref{eq:J.Eperp}): (a) $\textbf{J}\cdot \textbf{E}_{\perp,v}$, (b) $-\textbf{J}\cdot(\textbf{v}\times\textbf{B})$, (c) $(\eta/\Gamma)\textbf{J}^2$. The values are calculated at time $(t/\tau)_{20}$. The inset boxes in panel (a) show the domain considered to calculate flux: $x=$ 0.1 (green), 0.05 (blue), 0.025 (brown) from Table~\ref{tab:Erg_partition}. We overplot contours of the in-plane magnetic field streamlines to highlight the magnetic field structure.}
\end{figure}

Fig.~\ref{fig:Conv_res_spatial}(a) illustrates the spatial distribution of $\textbf{J}\cdot \textbf{E}$, quantifying energy conversion. The maximum energy transfer from electromagnetic field to the plasma occurs near the separatrix region, located just below the magnetic island. This region serves as a key site for energy deposition into the plasma as reconnection progresses. Most of the energy gained by the plasma is concentrated within the current sheet and the lower parts of the magnetic island along the separatrix near the reconnection region. Conversely, around the upper region of the magnetic island, the plasma loses energy back to the electromagnetic fields, suggesting a localized redistribution of energy during the evolution of the magnetic island.

\begin{figure}
\includegraphics[scale=0.25]{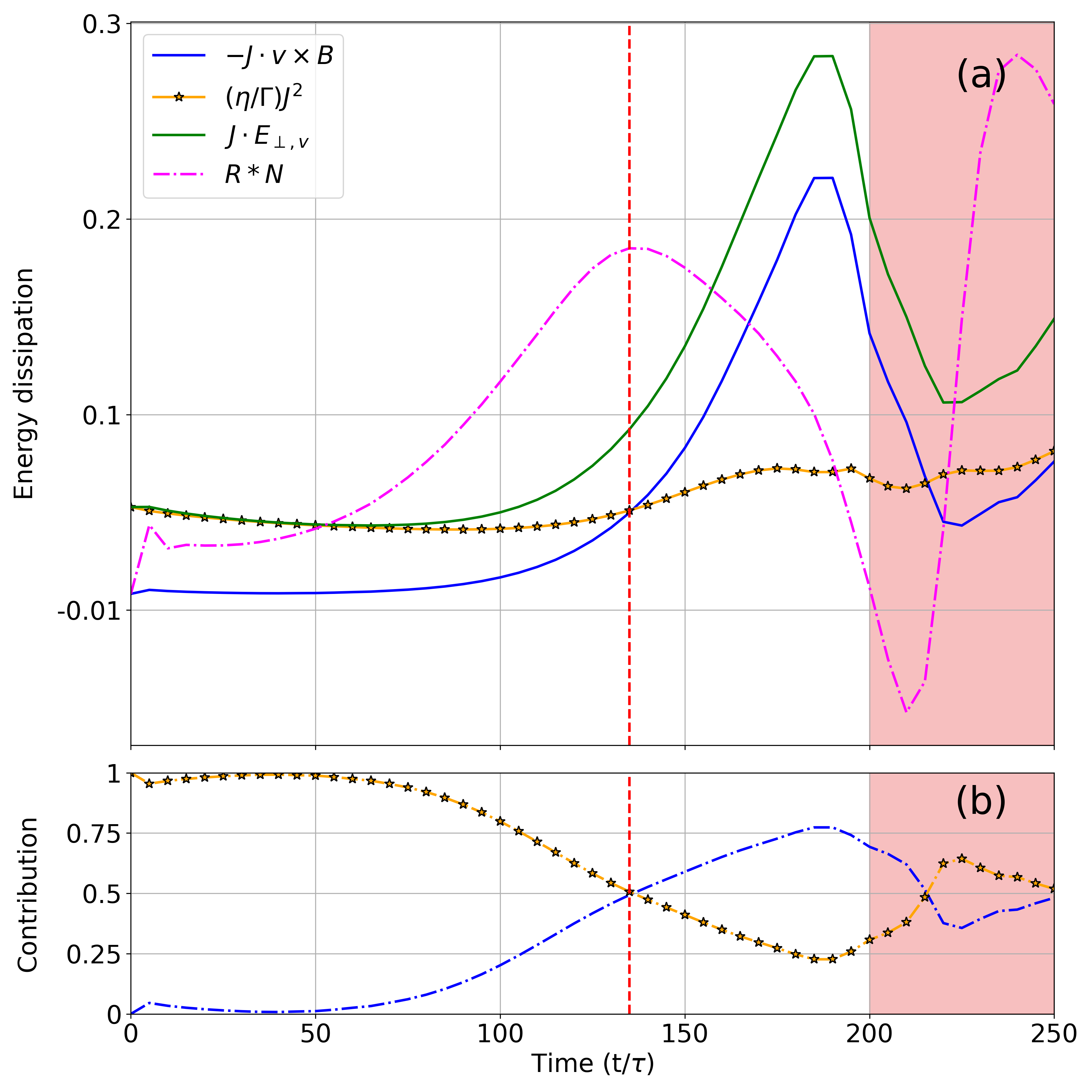}
\caption{\label{fig:Conv_res}Panel (a) shows the temporal analysis of terms of Eq.~(\ref{eq:J.Eperp}), where the energy conversion term (green) perpendicular to velocity field is decomposed into convective electric field (orange) and resistive electric field (blue), the reconnection rate (pink) is scaled by $N=12.5$ to visualize it on the same scales as the terms in Eq.~(\ref{eq:J.Eperp}) and refer the stages of reconnection with energy conversion evolution. Panel (b) shows the relative contribution which indicates that the resistive electric field dominates the initial stage of energy conversion, but as reconnection progresses, the contribution from the convective electric field increases. Here, the legends are the same as panel (a). During the peak reconnection phase shown by a vertical red dashed line, both fields contribute equally to the energy conversion. }
\end{figure}

In Fig.~\ref{fig:Conv_res} we analyze individual terms in Eq.~(\ref{eq:J.Eperp}) which quantify the contributions to the energy exchange process. The first term on the RHS represents the convective electric field ($-\textbf{v}\times \textbf{B}$) while the second term corresponds to the resistive electric field ($\eta \textbf{J}$). The resistive part $(\eta/\Gamma)\text{J}^2$ is initially ($t/\tau\leq135$) responsible for most of the energy conversion process. This behavior is expected, as resistivity facilitates magnetic reconnection by breaking the frozen-in condition. During this early phase, most of the electromagnetic energy is transferred to the plasma through resistive dissipation. Fig.~\ref{fig:Conv_res_spatial}(c) shows that this term only converts magnetic energy into plasma energy and not vice-versa, through resistive dissipation. This term is strongest within the current sheet and near the separatrix of the magnetic island, where resistivity facilitates the breaking of magnetic field lines and drives reconnection. \\
As the reconnection progresses the convective electric field begins to play a significant role. This term reflects the energy transfer associated with the bulk motion of the plasma as it accelerates out of the reconnection region. Over time ($t/\tau\geq135$), the convective electric field surpasses the resistive electric field in driving energy conversion. This transition marks a shift in the dominant energy transfer mechanism, as the plasma dynamics take precedence over resistive dissipation in the energy budget of the system. Fig.~\ref{fig:Conv_res_spatial}(b) shows that this term is both positive and negative in different regions, indicating its dual role in facilitating energy conversion. In regions near the sheet edge and around the magnetic island close to the sheet, this term converts electromagnetic energy into plasma energy whereas on the other end of the magnetic island, it converts plasma energy back into electromagnetic energy, highlighting its dynamic contribution to the energy exchange process.\\
At the time $t/\tau=135$, when the reconnection rate reaches its peak, the contributions from the convective and resistive electric fields to energy conversion become comparable. This represents a stage during the reconnection process, where both mechanisms work equally to sustain the energy transfer from the electromagnetic field to the plasma.

\subsection{\label{sigma-scale}Sigma Scaling}
\begin{figure}
\includegraphics[scale=0.25]{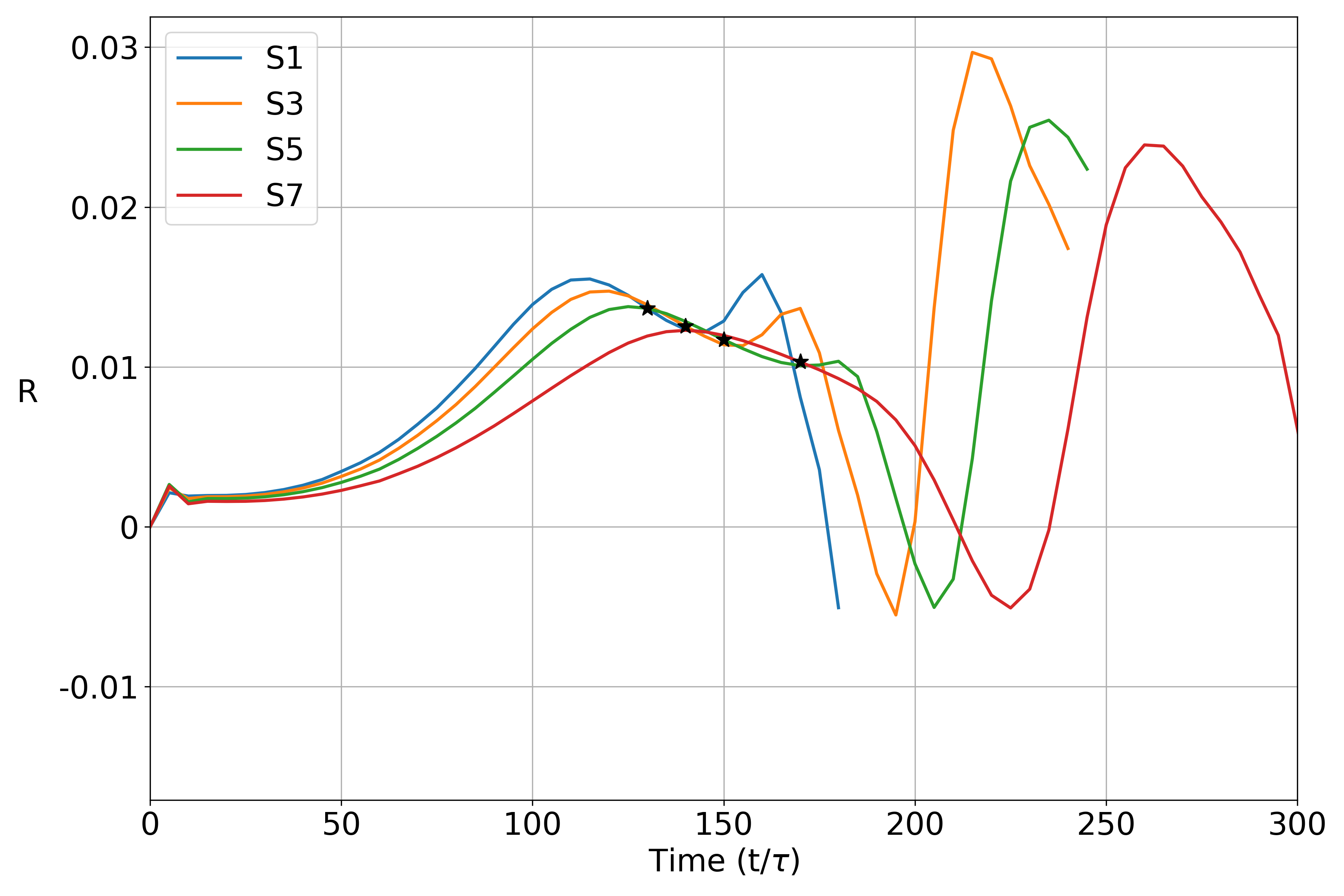}
\caption{\label{fig:Sigma_scan_all} Evolution of reconnection rate with time for selected $\sigma$ values from Table ~\ref{tab:sigma_scan}. Black markers indicate the $(t/\tau)_{20}$ time. }
\end{figure}

\begin{figure*}
\includegraphics[scale=0.25]{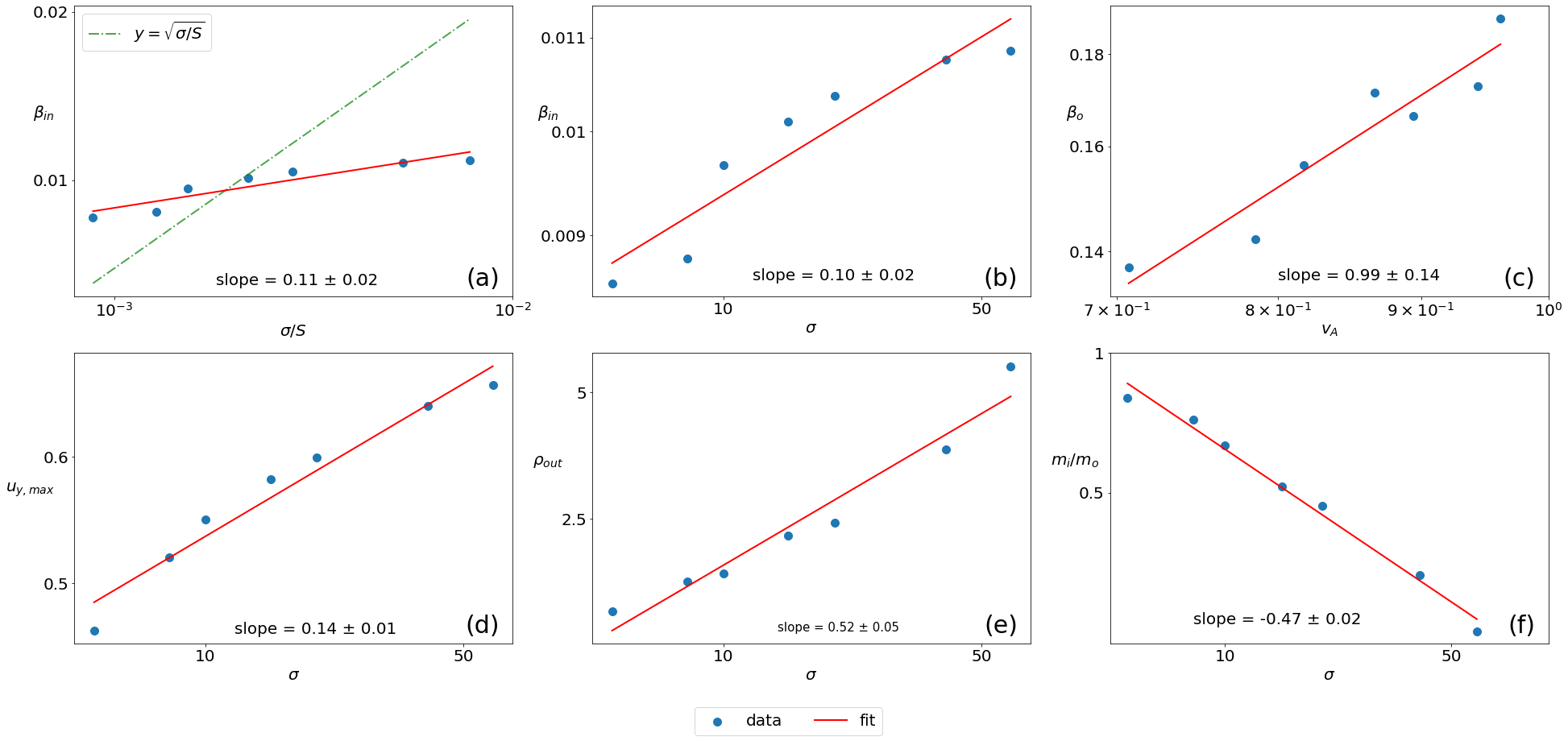}
\caption{\label{fig: comp_fact} Scaling analysis carried out for various parameters. Blue points indicate the value at $(t/\tau)_{20}$ time, and red lines are the curve-fit, whose slopes give the scaling. Panel (a) shows a weak scaling of $\beta_{in}$ with $\sigma/S$ versus the predicted scaling\cite{Lyutikov_2003} (green). Panels (b, e) show the scaling of $\beta_{in}$, $\rho_{out}$ with $\sigma$. Panel (c) shows a linear scaling of $\beta_{out}$ with $v_A$. Using these relations, we obtain a very strong scaling of compressibility factor $\alpha=m_i/m_o$ with $\sigma$ as shown in panel (f). The scaling in panel (d) for $u_{y,max}$ with $\sigma$ differs from the predicted value\cite{Lyubarsky05}.}
\end{figure*}

In [\onlinecite{Lyutikov_2003}] it was predicted that for a non-relativistic inflow i.e., when $\sigma\ll S$ and  $\beta_{in}\ll 1$, where $\beta=v/c$, the inflow velocity scales as $\beta_{in}\propto \sqrt{\sigma/S}$. In this regime $v_A\rightarrow c$, but resistivity is very small, so $S\gg \sigma$. This prediction assumes that magnetic energy is almost entirely spent on particle acceleration, with the plasma pressure being negligible compared to the bulk kinetic energy of the outflow. Lyubarsky\cite{Lyubarsky05} considered hot outflow and predicted a scaling for outflow as $\Gamma_{out}\sim\sqrt{\sigma}$.\\
We are studying scaling relations for mildly relativistic plasma of $\sigma$ in the range $(1-60)$. Our simulations are tabulated in Table~\ref{tab:sigma_scan} which analyzes the scaling behavior by studying the variation of inflow and ouflow properties with time for different values of $\sigma$. We have considered the blue domain ($0 \leq x\leq 0.05,\ 0 \leq y\leq 0.2$) in Fig.~\ref{fig:Conv_res_spatial}(a) for doing all the analysis in this segment, as it is close to the reconnection region and excludes the magnetic island. Fig.~\ref{fig:Sigma_scan_all} shows the temporal evolution of the inflow velocity ($\beta_{in}=v_x$, calculated similar to reconnection rate by taking an average over a line at $x=\pm0.05$ from $y=-0.2$ to $0.2$) for various $\sigma-$values, with the marker indicating the specific time when $\beta_{in}$ is measured. However, the plot of $\beta_{in}$ versus $\sigma/S$ in Fig.~\ref{fig: comp_fact}(a)  deviates from the predicted scaling\cite{Lyutikov_2003}, showing a much slower reconnection rate. We observe a weak scaling relation of $\beta_{in}\propto (\sigma/S)^{0.11\pm 0.02}$.
\\
\begin{table}
\caption{\label{tab:sigma_scan}
 Parameters for Sigma scan($R_M=800$)}
\begin{ruledtabular}
\begin{tabular}{ccccccc}
ID&$\sigma$&$\sigma_H$&$\beta_{m}$&$v_A$&$\beta_{in}(10^{-2})$&($t/\tau)_{20} $\\
\hline
S1& 60&  12& 0.03& 0.961& 1.08& 130\\
S2& 40&   8& 0.05& 0.943& 1.076& 135 \\
S3& 20&   4&  0.1& 0.894& 1.037& 140  \\
S4& 15&   3& 0.13& 0.866&  1.01& 145 \\
S5& 10&   2&  0.2& 0.817& 0.966& 150 \\
S6& 8& 1.6& 0.25& 0.785& 0.879& 155 \\
S7& 5&   1&  0.4& 0.707& 0.857& 170 \\
S8& 2& 0.4&  1.0& 0.535& 0.608& 225 \\
\end{tabular}
\end{ruledtabular}
\end{table}

Fig.~\ref{fig: comp_fact}(b) reveals that the inflow velocity experiences only a slight acceleration with increasing $\sigma$, ($v_A \sim c$) following the relation $\beta_{in}\propto \sigma^{0.1\pm 0.02}$. To better understand this weak dependence, we revisit the mass conservation equation:
\begin{equation}
    \beta_{in}=\rho_{out}\beta_{out}\delta/L,
    \label{eq: mass_cons}
\end{equation}

where $(\rho_{in}\approx 1$ and $\delta/L\approx B_{out}/B_{in}$ from $\nabla\cdot B=0$). We verify the ratio $\delta/L$, where $\delta$ is the late-stage sheet width and $L$ is the length of the elongated current sheet. The edges of the current sheet are identified as the locations where the current density drops to half of its peak value at the center. We find that this ratio remains nearly constant across all the $\sigma-$values considered. We explored the dependence of $\delta/L$ (or equivalently $B_{out}/B_{in}$) on $\sigma$ and saw that there was no clear scaling relation between $B_{out}/B_{in}$ and $\sigma$, instead the data points scatter around $\delta/L=0.05$, which was also verified from the geometry of the elongated sheet formed, suggesting that the reconnection layer thickness remains relatively constant across the $\sigma-$range around the $(t/\tau)_{20}$ time.
 The Alfven velocity calculated as $v_A=\sqrt{\frac{\sigma_H}{\sigma_H + 1}}$, is nearly independent of $\sigma$ for large $\sigma-$values. The scaling of $\beta_{out} \sim v_A^{0.99\pm 0.14} $ in Fig.~\ref{fig: comp_fact}(c) implies that even $\beta_{out} \sim v_A$ and hence it is independent of $\sigma$ at high $\sigma-$values. Outflow velocity ($\beta_{out}=v_y$) is calculated as the average flow velocity at the $L_{20}$ edge.
 Additionally, Fig.~\ref{fig: comp_fact}(e) reveals a scaling relation of $\rho_{out}\sim\sigma^{0.52\pm 0.05}$, indicating that the plasma density in the outflow increases with $\sigma$. We checked how the mass confined in the domain changes by integrating the density over the volume. We found that the ratio of inflow mass to outflow mass reduced to less than $40\%$, highlighting the effects of compressibility. We modified the mass conservation form to $\rho_{in}\beta_{in} L=\alpha \rho_{out}\beta_{out}\delta$, where $\alpha=Mass_{in}/Mass_{out}$ is the compressibility factor defined as the ratio between the inflow mass along the edge $L$ and outflow mass along the edge $\delta$, so we get $\beta_{in}\propto\alpha\rho_{out}$. That is, our inflow velocity is only dependent on the mass compression and the outflow density. Using the observed scaling relations in Fig.~\ref{fig: comp_fact}, we predict  $\alpha\sim\sigma^{(0.1\pm0.02)-(0.52\pm0.05)=-0.42\pm0.07}$. This is very well validated by Fig.~\ref{fig: comp_fact}(f) which shows a strong scaling of $\alpha\sim\sigma^{-0.47\pm 0.02}$, close to what we expected, highlighting the significance of the compressibility factor for different magnetizations. Fig.~\ref{fig: comp_fact}(d) illustrates a scaling relation for the maximum outflow speed $u_{y,max}\sim\sigma^{0.15\pm0.02}$ which is different from the prediction\cite{Lyubarsky05} that $\Gamma v_y \sim \sqrt{\sigma}$. 
\subsection{\label{Erg-Part}Energy Partition}
To investigate energy partitioning near the reconnection region, it is essential to analyze the different forms of energy flux involved in the process. In our study, we consider kinetic energy flux ($K_{i/o}$), magnetic energy flux ($B_{i/o}$), and thermal energy flux ($Th_{i/o}$) integrated along the edges of the domain under consideration, where $i/o$ represents inflow/outflow. These fluxes are defined as $\Gamma^2 \beta \rho L, \quad \beta B^2 L$ and $P\beta\Gamma^2L\gamma/(\gamma-1)$ respectively, where $\beta=v/c$ represents inflow and outflow speed, $L$ is the length of the edge of the domain under consideration. These flux definitions comes from the energy flux density, given by 

\begin{equation}
    \textbf{S}=\textbf{E}\times \textbf{B} + \rho\Gamma^2\textbf{v}\Big[1+\frac{\gamma P}{(\gamma-1)\rho}\Big],
    \label{eq: Flux_term_calc}
\end{equation}
It is clear from this expression that the second and third term on the RHS correspond to kinetic and thermal energy respectively. Let's look into the first term, here \textbf{E} is defined from relativistic Ohm's law as $\textbf{E}=\textbf{B}\times \textbf{v}+ \eta \textbf{J}/\Gamma$. The second term is very small at the edges and can be neglected, which gives us $\textbf{E}\times\textbf{B}\approx B^2\textbf{v}-(\textbf{v}\cdot\textbf{B})\textbf{B}$.

\begin{equation}
\begin{aligned}
    (\textbf{E}\times \textbf{B})_x=B_y^2\textbf{v}_x-\text{v}_y\text{B}_y\text{B}_x,\\
    (\textbf{E}\times \textbf{B})_y=B_x^2\textbf{v}_y-\text{v}_x\text{B}_y\text{B}_x,
\end{aligned}
\end{equation}
In the influx equation $(\textbf{E}\times\textbf{B})_x$ the second term is negligible as compared to first term, whereas in the outflux equation $(\textbf{E}\times\textbf{B})_y$ the second and first term are comparable but the overall contribution of magnetic energy to the energy partition in the outflow is insignificant, hence this can also be neglected. From this, we obtain the magnetic energy flux term as $B^2\beta L$ .\\
For this analysis, we focus on simulation S5. To examine the consistency of energy partition across different distances from the reconnection region, the analysis was repeated for different domain sizes. The top boundary ($y=0.2$), bottom boundary ($y=0.0$) and left boundary ($x=0.0$) were fixed, while the right boundary was varied with $x=[0.025,\ 0.05,\ 0.1]$ as shown in Fig.~\ref{fig:Conv_res_spatial}(a) in green, blue and brown boxes respectively. The domain was kept close to the reconnection region, ensuring that the magnetic island was excluded from the analysis. \\
\begin{table}
\caption{\label{tab:Erg_partition}
 Energy partition at ($t/\tau=155, \quad\sigma=10)$}
\begin{ruledtabular}
\begin{tabular}{c|cccc|cccc}
D(x)&$K_{i}$&$B_{i}$&$Th_{i}$&$\sum_{i}$&$K_{o}$&$B_{o}$&$Th_{o}$&$\sum_{o}$\\
\hline
0.025&12.65&104.16&50.2 &167&22.29&0.36&164.4 &187 \\
0.05 &13.09&111.55&52.03&177&22.63&0.4 &165.78&189  \\
0.1  &12.32&109.12&49.14&171&20.98&0.29&159.21&180 \\
\hline
Mean &12.69&108.28&50.46&171&21.97&0.35&163.1&185 \\
\end{tabular}
{\raggedright \textbf{Notes:} Column 1: D refers to the domain where three edges are fixed at $y=0,\ y=0.2,\ x=0 $ and the position of the fourth edge is varied; Row 1: $i/ o$ refers to inflow/ outflow flux of energies.\par}
\end{ruledtabular}
\end{table}

From Table~\ref{tab:Erg_partition}, it is evident that the energy partition remains consistent across different domain sizes as the mean values are close to the values of different domain sizes. The total energy at the inflow and at the outflow boundaries are nearly equal, with a difference of only about $8\%$. The inflow is dominated by the magnetic energy which constitutes approximately $63\%$ of total energy flux, followed by the energy flux due to thermal pressure which accounts for around $30\%$, while kinetic energy makes up a very small portion. The outflow is almost entirely made up of thermal and kinetic flux at $88\%$ and $11.9\%$ respectively. Magnetic flux plays a negligible role. A schematic of this is presented in Section \ref{Guide-field}. This indicates that most of the incoming magnetic energy goes to plasma heating, highlighting the significant role of thermal heating during the reconnection process\cite{Zenitani_2010}. This finding is consistent with previous results from particle-in-cell (PIC) simulations\cite{Mbarek}.

\begin{figure}
\includegraphics[scale=0.25]{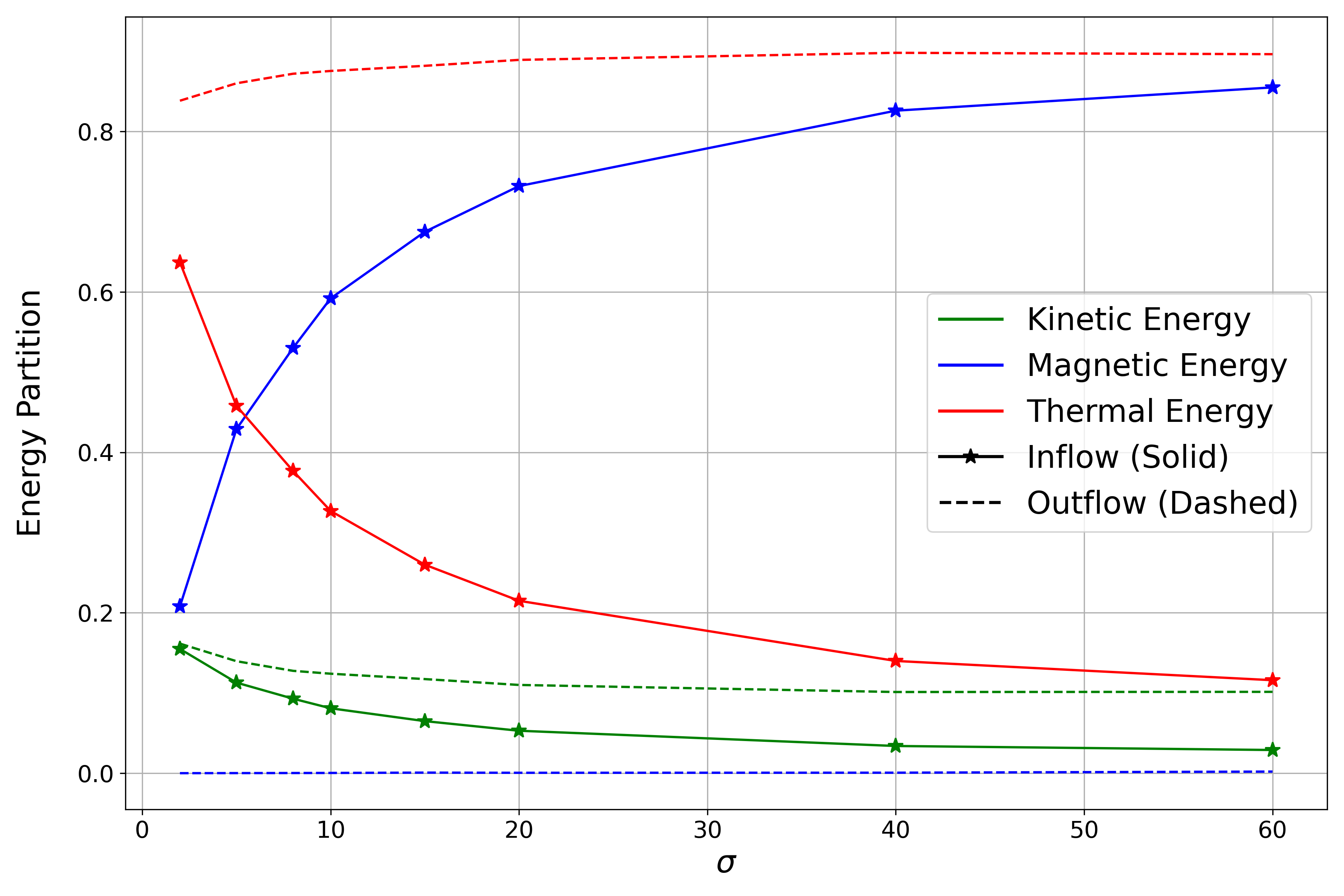}
\caption{\label{fig:Erg_sigma} Energy partition of kinetic energy(green), magnetic energy(blue) and thermal energy(red) in the inflows(solid) and ouflows(dashed). Values are calculated at $(t/\tau)_{20}$ for different $\sigma$, which shows the dominance of thermal energy in the outflow. }
\end{figure}
Fig.~\ref{fig:Erg_sigma} shows the variation in the energy composition of the inflow and outflow with different values of $\sigma$ for the simulations in Table ~\ref{tab:sigma_scan}. For the incoming flow, as $\sigma$ increases, the contribution of magnetic energy in the flux rises significantly, which accounts for a drop in the contribution of thermal energy. The outflow remains overwhelmingly dominated by thermal energy, which consistently accounts for approximately $90\%$ of the total energy flux even at higher $\sigma-$values. The kinetic energy contribution decreases slightly with increasing $\sigma$, but the overall composition of the outflow remains relatively unchanged. To make sure that the energy of the system remains conserved throughout the simulation we have measured and compared the change in the rate of energy and flux in Appendix \ref{Appendixes}.

Fig.~\ref{fig:Bsq_P_x_sigma} presents the variation of magnetic pressure and thermal pressure as a function of the distance from the centre of the current sheet for different $\sigma-$values. It shows that even though the magnetic field strength increases with larger $\sigma$, it fails to penetrate deeper into the sheet due to the dominance of thermal pressure. The balance point remains near the edge of the sheet even for large $\sigma-$values. The values were calculated at $(t/\tau)_{20}$ time.
The presence of a large thermal pressure term does not allow the reconnection rate to proceed at as faster rate as predicted\cite{Lyutikov_2003}.
\begin{figure}
\includegraphics[scale=0.25]{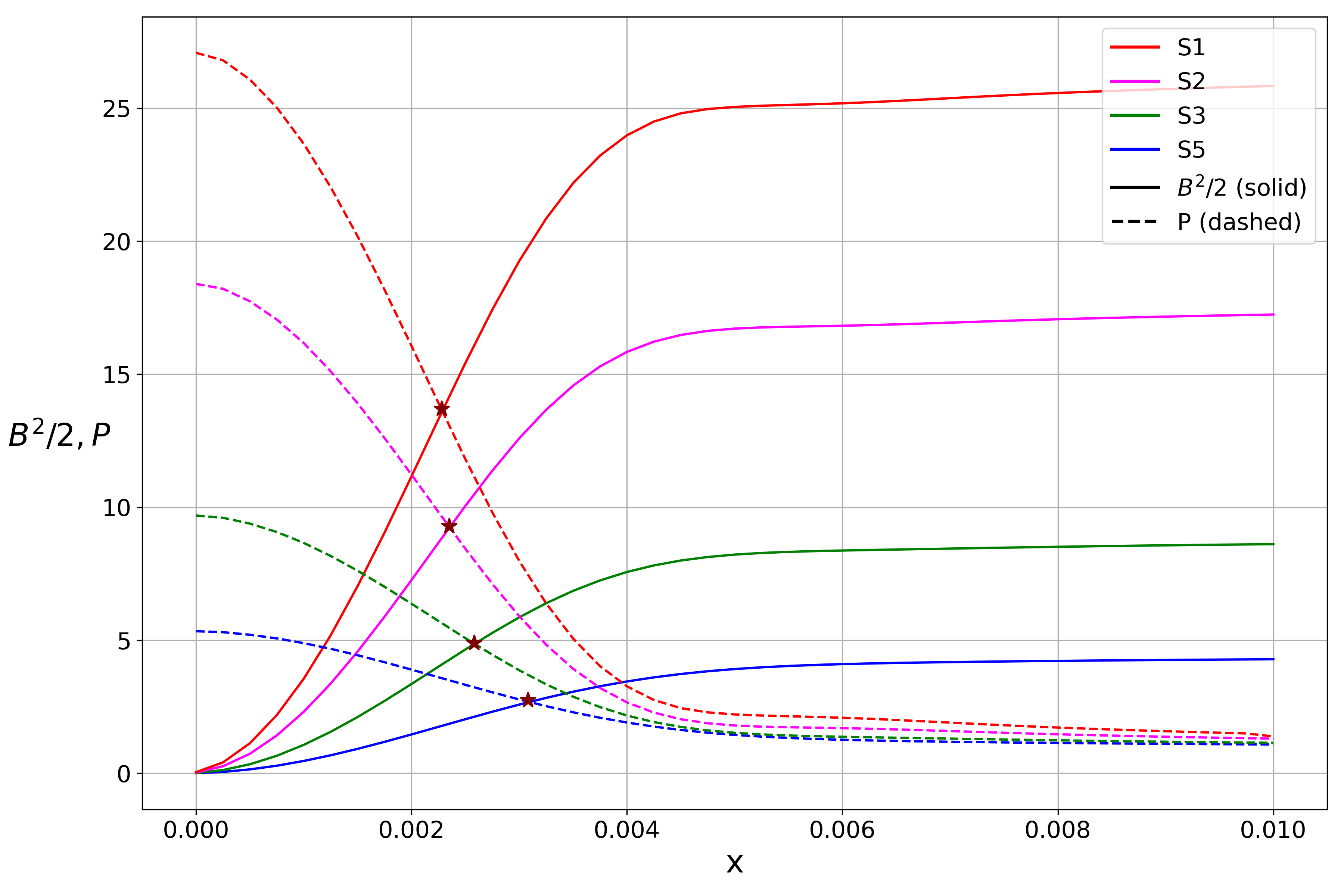}
\caption{\label{fig:Bsq_P_x_sigma} Variation of $B^2/2$ (solid) and P (dashed) near the vicinity of the current sheet for different values of $\sigma$ from Table~\ref{tab:sigma_scan}. The marker indicates the edge of the current sheet, where magnetic and thermal pressure balance is achieved, highlighting the inability of larger $\sigma$ to further thin out the sheet. The values are calculated at $(t/\tau)_{20}$.}
\end{figure}
\subsection{\label{Guide-field}Guide field}
We introduce a constant and uniform guide field ($B_G$) perpendicular to the antiparallel magnetic field to study its influence on the reconnection rate and energy partition. The modified magnetic field configuration is given by $$\textbf{B} = B_0\tanh(2x/\lambda)\hat{j} + B_G\hat{k},$$ where $B_G/B_0$ is varied as $[0,\ 0.25,\ 0.33,\ 0.5,\ 0.75,\ 1.0,\ 1.5]$. Table~\ref{tab: Recc_rate_Bg} shows the reconnection rate and time $(t/\tau)_{20}$ for different setups. Plotting the graph of reconnection rate at $(t/\tau)_{20}$ versus $B_0/B_G$ reveals that the reconnection rate decreases with increasing guide field\cite{FU2007117,Huba2005} and the time taken to reach $(t/\tau)_{20}$ increases. We do not see a clear $\sigma$ scaling across ($B_0/B_G$). Interestingly, in regimes where the guide field is strong ($B_G / B_0 \geq 0.75$), some hints of scaling behavior begin to emerge. Further investigation is required to understand better the reconnection dynamics in these stronger guide field environments and to find any underlying scaling laws that may govern this regime.\\
\begin{table}
\caption{\label{tab: Recc_rate_Bg}
 Parameters for guide field scaling(R$_m=500$)}
\begin{ruledtabular}
\begin{tabular}{cccc}
ID &$B_G/B_0$&Reconnection rate(R)$ (10^{-2})$&($t/\tau)_{20} $\\
\hline
G1& 0  & 1.5 & 155  \\
G2& 0.25 & 1.44 & 160  \\
G3& 0.33 & 1.39 & 165  \\
G4& 0.5  & 0.24  & 175   \\
G5& 0.75 & 0.93 & 190 \\
G6& 1 & 0.61 & 215  \\
G7& 1.5 & 0.36 & 275 \\
\end{tabular}
\end{ruledtabular}
\end{table}

\begin{figure}
\includegraphics[scale=0.5]{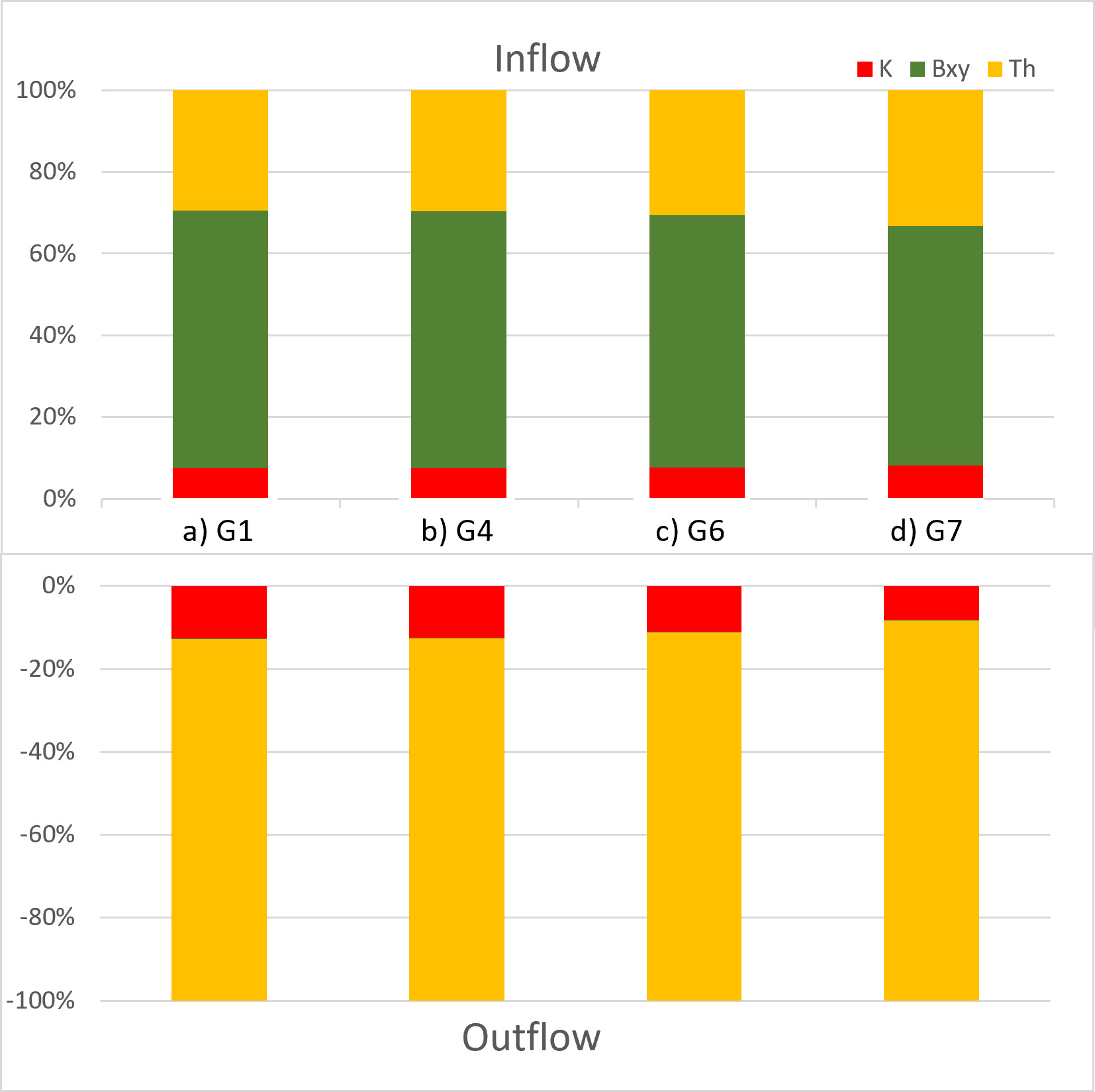}
\caption{\label{fig:Erg_Bg} Variation of energy partition with an increasing guide field, considering the in-plane magnetic field energy (green), kinetic energy (red), and thermal energy (orange), which hardly shows significant change with change in guide field. }
\end{figure}

Fig.~\ref{fig:Erg_Bg} shows how the energy partition varies in the inflow and outflow for a changing $B_G/B_0$ ratio. We find that, with increasing guide field strength relative to $B_0$, the energy contribution from the guide field itself becomes significant for both inflow and outflow regions which was also seen in [\onlinecite{Zenitani_2009}]. However, when the energy partition is recalculated by excluding the guide field's contribution, the relative distribution of energy among the kinetic, magnetic, and thermal components remains largely unchanged and shown in Fig.~\ref{fig:Erg_Bg}(a).\\
We also simulated reconnection in a current sheet by using a rotating guide field\cite{grehan2025}, where the magnetic field has the form $B=B_0\tanh(2x/\lambda)\hat{j} + B_0 \text{sech}(2x/\lambda)\hat{k}$. In this configuration, a pressure gradient is not required to maintain force balance, as the system is force-free, where $\textbf{J}\times \textbf{B}=0$. The variation in the guide field $(B_Z)$ provides the necessary current structure, and the plasma pressure remains uniform $(\nabla P=0)$. In our simulations, first, we kept the values of pressure and density uniform, and then decreased their values. However, these variations had no significant effect on the energy composition of the outflow. The thermal energy remained the dominant component, accounting for nearly $90\%$ of the total outflow energy. This shows that the internal pressure of the current sheet is not important in conversion to thermal energy.


\section{Conclusions}
In this study, we conducted special relativistic resistive magnetohydrodynamics simulations to investigate the dynamics of relativistic reconnection for collisional and macroscopic plasma environments. We observed how reconnection rates and current sheet structures evolve under varying plasma conditions and validated the Sweet-Parker scaling. For energy conversion analysis we considered the relativistic Ohm's law and decomposed the energy conversion term into resistive and convective part. We found that both terms contribute equally to the energy conversion process when the reconnection rate peaks. The maximum energy conversion occurs near the separatrix of the magnetic island close to the current sheet, where the plasma gains maximum energy from the electromagnetic field and dissipates part of it at the opposite end of the island. The convective electric field plays an important role in both these processes, while, the resistive part stays concentrated within the current sheet, dominating the initial phase of energy conversion. This is similar to the results obtained in \cite{French_2023} where initially, resistive electric field does most of the work.

The weak $\sigma-$scaling observed in both the inflow and outflow regions, along with the results from our energy partition analysis, indicates a deviation from previously predicted scaling laws. Specifically, we find that the inflow follows a scaling of $\beta_{in} \propto (\sigma/S)^{0.11}$, which is significantly weaker than the prediction\cite{Lyutikov_2003} $\beta_{in}\propto(\sigma/S)^{0.5}$, derived by assuming most of the energy getting converted to kinetic energy. However, our simulation shows that most of the energy is transferred to thermal energy, which was also seen in PIC simulation\cite{Mbarek}. Similarly, the outflow exhibits a scaling of $u_{y,max}\propto\sigma^{0.15}$ which is weaker than the prediction of $\Gamma v_y\sim\sigma^{0.5}$ (where, $u=\Gamma v$), derived assuming incompressibility. However, our simulation shows strong compression with good scaling, and also the mass inflow and outflow are not in true steady state in the Sweet-Parker regime. Interestingly, in the plasmoid dominated nonlinear regime Grehan et al.,\cite{grehan2025} has verified the predicted scaling. The difference underscores the critical importance of accounting for compressibility and thermal energy in both inflows and outflows during the magnetic reconnection process. It has also been proposed\cite{Parker1963} that reconnection rate should increase by a factor of $\sqrt{\rho_m/\rho_0}$, where $\rho_m$ is the density within the current sheet and $\rho_0$ is the density in the ambient plasma, but we do not observe this in our simulation. Furthermore, the introduction of an external out-of-plane guide field provided insights into the dependence of the reconnection rate on the guide field. The results consistently showed that outflow regions remain strongly dominated by thermal energy, regardless of the guide field strength. Even in scenarios where we implemented a rotating guide field with initially low pressure, the thermal pressure continued to dominate the outflow energy budget. Global general relativistic resistive MHD simulations showed that magnetic reconnection occurs near the event horizon, where it can result in energy release that can be observed in flares of radiation\cite{Bart2020}. These simulations emphasize that thermal effects and compressibility are important for the dynamics of reconnection. Our study gives a scaling for compressibility and thermal effects in the collisional regime. This is crucial for developing a more comprehensive understanding of energy conversion mechanisms in high-energy astrophysical plasmas. 
 
\newpage

\begin{acknowledgments}
HP and KM would like to thank DST – FIST (SR/FST/PSI-215/2016) for the financial support alongwith Indian Institute of Technology, Hyderabad seed grant. The support and the resources provided by the “PARAM Seva facility” under the National Supercomputing Mission, Government of India at the Indian Institute of Technology, Hyderabad are gratefully acknowledged.\\
BR acknowledges support by the Natural Sciences \& Engineering Research Council of Canada (NSERC), the Canadian Space Agency (23JWGO2A01), and by a grant from the Simons Foundation (MP-SCMPS-00001470). BR acknowledges a guest researcher position at the Flatiron Institute, supported by the Simons Foundation.

\end{acknowledgments}

\section*{Data Availability Statement}
The data that support the findings of this study are available from the corresponding author upon reasonable request.

\appendix

\section{\label{Appendixes}Energy Conservation}
\begin{figure}
\includegraphics[scale=0.25]{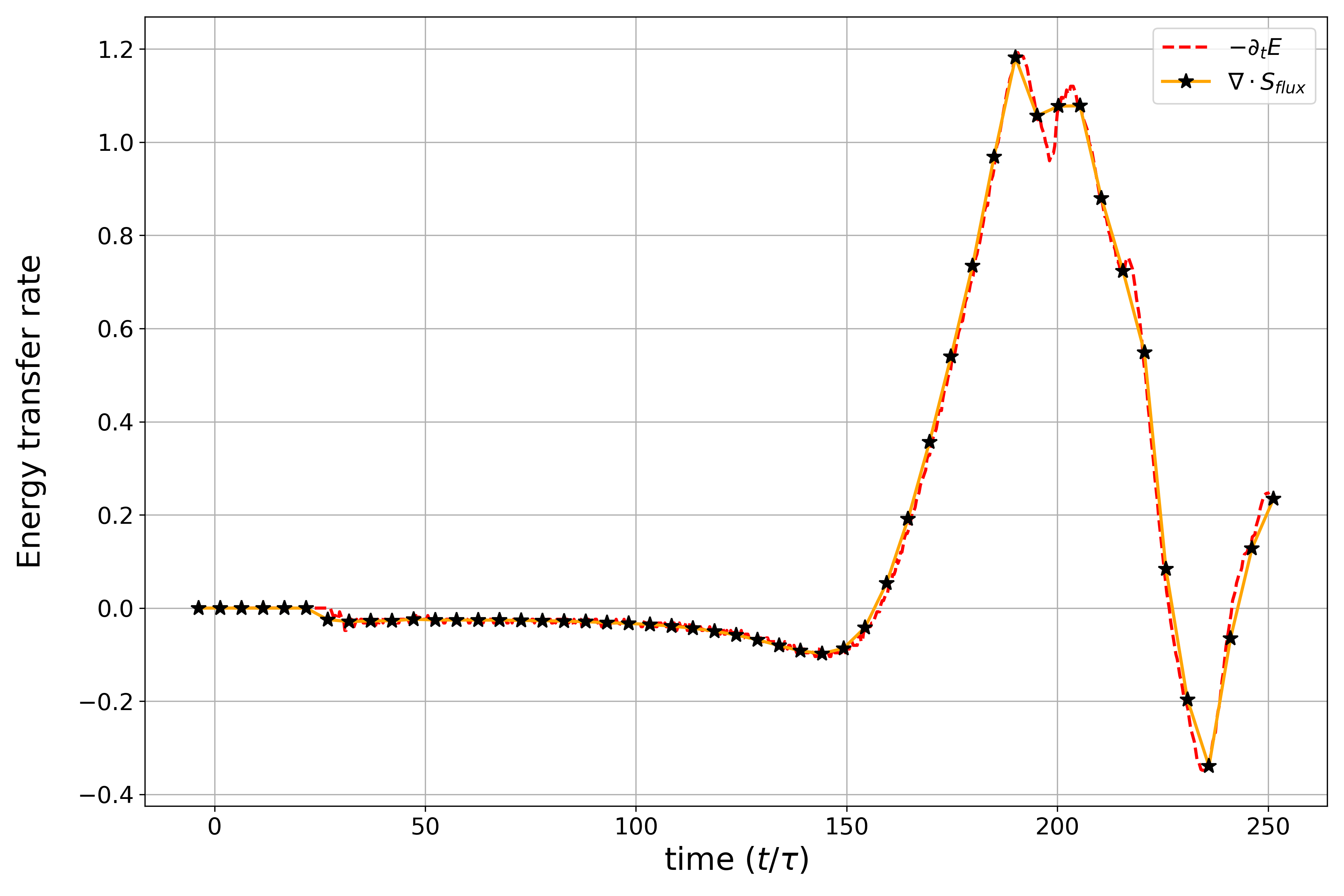}
\caption{\label{fig:Erg_conserv} Energy conservation in the simulation is shown by matching the rate of change of energy density (red) to the flux across the boundaries (yellow) plotted over the simulation time. The above analysis is for the S5 model.}
\end{figure}

To ensure that total energy is conserved throughout our simulations, we conducted a comprehensive energy conservation analysis for our simulation models. Since, we employed continuous boundary conditions, it was essential to verify the rate of change of energy within the simulation domain matches with the flux of energy through the boundaries. This helps us to confirm that no artificial energy gain or loss occurs within the system due to numerical artifacts or boundary effects. The energy conservation equation is given by
\begin{equation}
   \frac{\partial \tau}{\partial t} + \nabla\cdot \textbf{S} = 0 
\end{equation}
where $\tau$ represents the total energy density, and $\textbf{S}$ is the energy flux vector. To compute the total energy density ($\tau$) within the domain we considered contribution from kinetic energy ($\rho\Gamma^2$), electromagnetic energy ($(E^2 + B^2)/2$) and thermal energy ($4P\Gamma^2 -P$). To calculate the energy flux, we integrate $\textbf{S}=\textbf{E}\times\textbf{B} + \rho h\Gamma^2\textbf{v}$ on the boundary. The first term in the flux equation is associated with the electromagnetic energy transfer while the second term accounts for the flux of kinetic and thermal energy. Here, $h=1+\frac{\gamma P}{(\gamma -1)\rho}$ is specific enthalpy. Fig.~\ref{fig:Erg_conserv} shows the comparison between the time derivative of the total energy density within the simulation domain and the net energy flux across the boundaries for the S5 simulation. The two quantities are nearly equal throughout the simulation (for $dt=0.3\ t/\tau$) with a maximum difference of $10\%$ near time $(t/\tau=190)$ far from $(t/\tau)_{20}=155$  indicating that our simulation conserves the total energy satisfactorily, validating both the numerical methods and boundary conditions used in our model.

\nocite{*}
\bibliography{aipsamp}

\end{document}